# Small-scale Flux Emergence, Coronal Hole Heating, and Flux-tube Expansion: A Hybrid Solar Wind Model

Y.-M. Wang
Space Science Division, Naval Research Laboratory, Washington, DC 20375-5352, USA; yi.wang@nrl.navy.mil

## Abstract

Extreme-ultraviolet images from the Solar Dynamics Observatory often show loop-like fine structure to be present where no minority-polarity flux is visible in magnetograms, suggesting that the rate of ephemeral region (ER) emergence inside "unipolar" regions has been underestimated. Assuming that this rate is the same inside coronal holes as in the quiet Sun, we show that interchange reconnection between ERs and open field lines gives rise to a solar wind energy flux that exceeds $10^5$ erg cm$^{-2}$ s$^{-1}$ and that scales as the field strength at the coronal base, consistent with observations. In addition to providing ohmic heating in the low corona, these reconnection events may be a source of Alfvén waves with periods ranging from the granular timescale of ∼10 minutes to the supergranular/plume timescale of many hours, with some of the longer-period waves being reflected and dissipated in the outer corona. The asymptotic wind speed depends on the radial distribution of the heating, which is largely controlled by the rate of flux-tube expansion. Along the rapidly diverging flux tubes associated with slow wind, heating is concentrated well inside the sonic point (1) because the outward conductive heat-flux density and thus the outer coronal temperatures are reduced, and (2) because the net wave energy flux is dissipated at a rate proportional to the local Alfvén speed. In this "hybrid" solar wind model, reconnection heats the lower corona and drives the mass flux, whereas waves impart energy and momentum to the outflow at greater distances.

*Unified Astronomy Thesaurus concepts:* Solar coronal holes (1484); Solar coronal heating (1989); Solar corona (1483); Solar coronal plumes (2039); Solar wind (1534); Slow solar wind (1873); Fast solar wind (1872); Solar magnetic fields (1503); Solar magnetic flux emergence (2000); Solar magnetic reconnection (1504); Alfven waves (23); Heliosphere (711)

## 1. Introduction

Models for coronal hole heating and solar wind acceleration have conventionally been divided into two kinds: those based on the generation and subsequent dissipation of magnetohydrodynamic (MHD) waves, and those invoking magnetic reconnection. In the first type of model, granular and supergranular convective motions continually jostle the embedded open field lines, exciting transverse oscillations in the form of Alfvén waves. Although most of these waves are trapped because of the steep density falloff in the chromospheric–coronal transition region, a small fraction leaks out into the corona; some of the outgoing waves undergo reflection due to the radial gradients in the Alfvén speed, $v_A$, leading to nonlinear interactions and the dissipation of energy via a turbulent cascade (see, e.g., Heinemann & Olbert 1980; Hollweg 1986; Moore et al. 1991; Velli 1993; Dmitruk et al. 2001, 2002; Cranmer & van Ballegooijen 2005; Verdini et al. 2005, 2010, 2019; Cranmer et al. 2007; Cranmer 2010; van Ballegooijen & Asgari-Targhi 2016). Gradients in the Alfvén wave pressure may also impart momentum to the wind plasma, a process that may be especially effective beyond the sonic point (Leer & Holzer 1980).

In the second type of model, the open field lines undergo footpoint exchanges or "interchange" reconnection with small loops that emerge inside the coronal hole, injecting energy into the corona in the form of ohmic dissipation and plasma jets (Parker 1991; Axford et al. 1999; Fisk et al. 1999; Cranmer & van Ballegooijen 2010). The two scenarios are by no means mutually exclusive. Reconnection events are likely to be an important source of MHD waves, while photospheric convective motions will drive interchange reconnection as well as directly launching waves along open field lines (see, e.g., Cranmer 2018). Indeed, both reconnection and wave generation may be needed to account for coronal hole heating and solar wind acceleration. Reconnection around null points and separatrices in the low corona will result in strong ohmic heating, raising the temperature and helping to drive the mass flux from below. Alfvén waves, either originating from such reconnection events or leaking out from below, will propagate outward to greater distances before undergoing reflection and turbulent dissipation, helping to accelerate the wind rather than boosting the mass flux.

A strong argument against a major role for reconnection in coronal hole heating has been raised by the statistical study of ephemeral regions (ERs) by Hagenaar et al. (2008, 2010). They found that the rate at which these small bipoles emerged inside coronal holes and other unipolar regions was at least a factor of 3 lower than in the quiet Sun. Based on these measurements, Cranmer & van Ballegooijen (2010) concluded that the rate of interchange reconnection inside coronal holes was too small to drive the solar wind. Moreover, statistical studies of X-ray jets inside polar coronal holes indicate that their total energy flux density is only ∼$10^3$ erg cm$^{-2}$ s$^{-1}$, two orders of magnitude lower than required for the solar wind (Sako et al. 2013; Paraschiv et al. 2015). However, there may be large numbers of smaller jets or "jetlets" that are not detected in X-rays (Tian et al. 2011; Raouafi & Stenborg 2014; Panesar et al. 2018).

A potential problem with the wave heating scenario is that Alfvén waves having wavelengths $\lambda < \lambda_{\rm crit} \sim 2\pi v_A |\partial v_A/\partial r|^{-1} \sim 2\pi R_\odot$, corresponding to periods $\lesssim 2\pi R_\odot/v_A \sim$ 1 hr, may undergo very little reflection near the Sun. If granular motions are the main driver of these waves, their periods would be typically on the order of 10 minutes. Wave reflection and dissipation might be enhanced by the presence of density inhomogeneities such as those associated with slow-mode waves, as in the model of Suzuki & Inutsuka (2006). In any case, however, the in situ solar wind is observed to be dominated by long-period Alfvén waves, whose origin has yet to be understood (see, e.g., Belcher & Davis 1971; Hollweg & Isenberg 2007; Roberts 2010).

Our purpose here is to suggest that the rate of small-scale flux emergence inside coronal holes has been underestimated, and that the energy input associated with interchange reconnection may be sufficient to drive the solar wind mass flux. In addition, the ubiquitous presence of diffuse plume emission inside coronal holes is interpreted as evidence for continual reconnection driven by supergranular flows, which in turn gives rise to Alfvén waves with periods ranging up to 1 day. Finally, we discuss how the rate of flux-tube expansion affects the radial distribution of coronal hole heating and thus the asymptotic wind speed.

## 2. The Emergence Rate of Ephemeral Regions inside Coronal Holes

Using line-of-sight magnetograms recorded with the Michelson Doppler Imager (MDI) during 2000–2005, Hagenaar et al. (2008, 2010) found that the rate of ER emergence depended on the flux imbalance parameter $\xi \equiv \Phi_{\rm net}/\Phi_{\rm abs}$, where $\Phi_{\rm net}$ ($\Phi_{\rm abs}$) represents the signed (unsigned) flux summed over a $92 \times 92$ Mm$^2$ area surrounding the ERs. The emergence rate $E_{\rm ER}$ was at least a factor of 3 smaller in strongly unipolar ($\xi \simeq 1$) regions than in more mixed ($\xi \lesssim 0.5$) regions. As fitted by Cranmer & van Ballegooijen (2010),

$$E_{\rm ER}(\xi) \simeq (92.9 - 68.5\xi^2) \text{ Mx cm}^{-2} \text{ day}^{-1}$$
$$\simeq (1.08 - 0.79\xi^2) \times 10^{-3} \text{ Mx cm}^{-2} \text{ s}^{-1}. \quad (1)$$

The measurements of Hagenaar et al. were based on 5 minute averages of magnetograms with 2″ pixels, with the individual ERs having total unsigned fluxes $\Phi_{\rm ER} \gtrsim 4 \times 10^{18}$ Mx. In an earlier study also employing MDI magnetograms, Abramenko et al. (2006) found that the rate of ER emergence inside coronal holes was a factor of $\sim$2 smaller than in the neighboring quiet Sun. As pointed out by Hagenaar et al. (2008), this reduction is a general property of unipolar regions, not just of coronal holes, contrary to the prediction of Fisk (2005).

Since 2010, the Solar Dynamics Observatory (SDO) has been providing $\sim$1″ resolution observations of the photospheric field and the extreme-ultraviolet (EUV) corona. In comparisons between Helioseismic and Magnetic Imager (HMI) line-of-sight magnetograms and Atmospheric Imaging Assembly (AIA) images taken in the 17.1, 19.3, and 21.1 nm bandpasses, we have identified numerous cases in which small coronal loops of horizontal size $\sim$2–5 Mm were embedded inside strong unipolar network or active region plages, but no underlying minority-polarity flux was visible in the corresponding magnetograms (Wang 2016a; Wang et al. 2019). Likewise, inside coronal holes, bright plume emission is often observed above network concentrations where very little minority-polarity flux is seen in magnetograms, but where EUV images show clusters of loop-like features (Wang et al. 2016); an example is displayed in Figure 1. The frequent absence of corresponding minority-polarity signals in the magnetograms is more likely to be the result of insufficient instrument sensitivity than of inadequate spatial resolution, since the horizontal extents of the loops greatly exceed the 0″.5

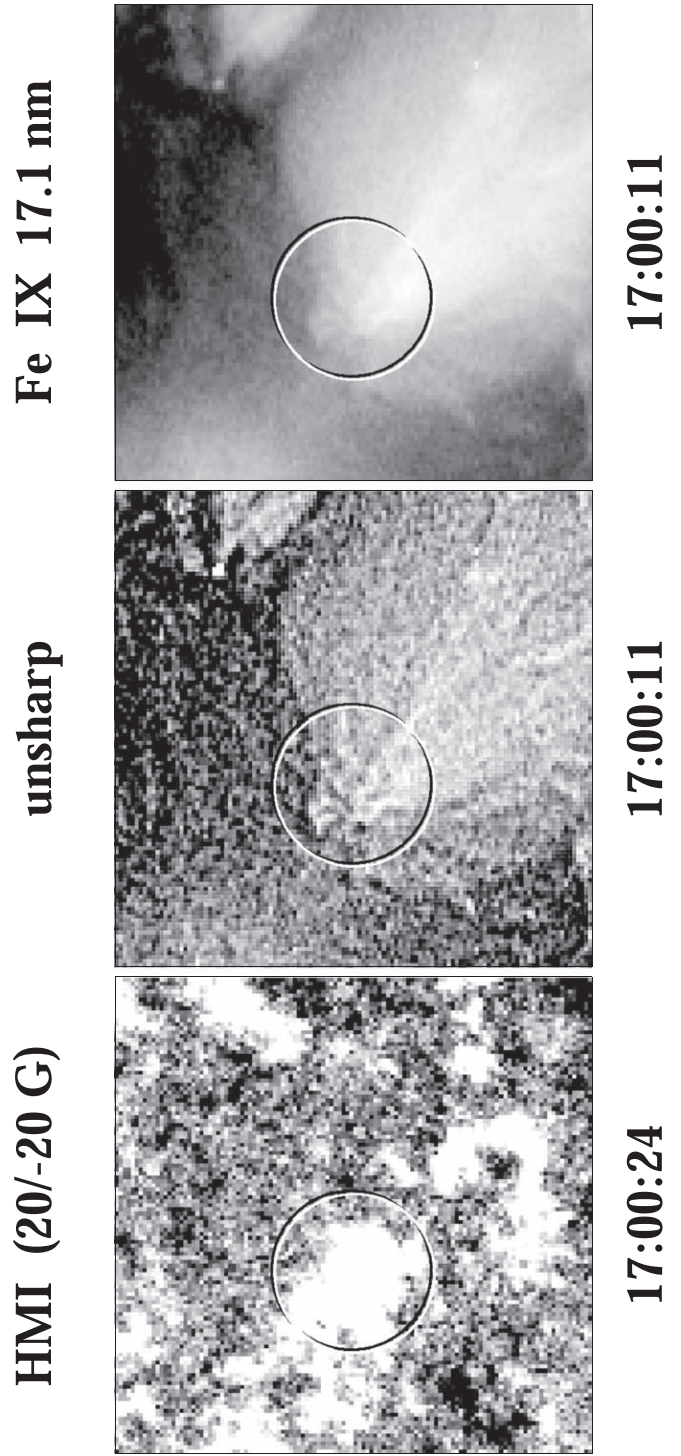

**Figure 1.** Coronal plume observed by SDO inside a northern hemisphere coronal hole on 2014 December 12. The core of the plume contains a collection of loop-like features, although little underlying minority-polarity flux is seen. Top panel: Fe IX 17.1 nm image recorded at 17:00:11 UTC. Middle panel: same, but with an unsharp mask applied. Bottom panel: HMI line-of-sight magnetogram taken at 17:00:24 UTC and saturated at ±20 G. Circled area has radius 12″; the enclosed loops have horizontal scale sizes of order 4 Mm. In the magnetogram, white (black) represents positive-polarity (negative-polarity) field.

HMI pixel size. We also note that minority flux that emerges within a strong unipolar region will tend to cancel and resubmerge on shorter timescales than it would if it emerged in a region of weaker or mixed field. The presence of a stronger background field at sunspot maximum might explain why Hagenaar et al. (2008, 2010) measured lower rates of ER emergence in 2000–2002 than in 2004–2005, independent of the flux imbalance parameter $\xi$ (see their Figure 5).

Because of line profile saturation and temperature effects, solar magnetograms tend to underestimate the actual photospheric field strengths, as deduced from interplanetary field measurements, by factors of ∼2–4 (see, e.g., Wang & Sheeley 1995; Ulrich et al. 2002; Linker et al. 2017). By cross-correlating MDI data with saturation-corrected magnetograms from the Mount Wilson Observatory (MWO), Tran et al. (2005) derived a correction factor of 1.7 for the MDI measurements.

We shall henceforth assume that the rate of ER emergence is the same inside coronal holes as in the quiet Sun. Setting $\xi = 0$ in Equation (1) and scaling the measured fluxes upward by 1.7, we obtain

$$E_{ER}(\Phi_{ER} \gtrsim 6.8 \times 10^{18} \text{ Mx}) \simeq 1.83 \times 10^{-3} \text{ Mx cm}^{-2} \text{ s}^{-1} \quad (2)$$

for the emergence rate of ERs having total unsigned fluxes $\Phi_{ER} \gtrsim 6.8 \times 10^{18}$ Mx.

The rate of small-scale flux emergence, in the form of ERs and intranetwork flux, is known to exhibit an exponential or power-law dependence that continues to increase toward smaller spatiotemporal scales (see, e.g., Thornton & Parnell 2011; Zhou et al. 2013). Because our focus is on reconnection events occurring in the corona, we include only the contribution of bipoles whose scale sizes exceed ∼2 Mm. According to Hagenaar (2001), an ER having a total flux $\Phi_{ER}$ of $1.13 \times 10^{19}$ Mx (or $1.92 \times 10^{19}$ Mx when corrected for line profile saturation) has an average pole separation $d_{ER}$ of 8.9 Mm. Assuming that $d_{ER} \propto \Phi_{ER}^{1/2}$ (following Cranmer & van Ballegooijen 2010) and taking the vertical scale size $h_{ER}$ of an ER to be comparable to its pole separation $d_{ER}$, we may write

$$h_{ER} \sim 2.0 \left( \frac{\Phi_{ER}}{1 \times 10^{18} \text{ Mx}} \right)^{1/2} \text{ Mm.} \quad (3)$$

This means that ERs with fluxes $\Phi_{ER} \gtrsim 10^{18}$ Mx will have loop systems that extend to heights above ∼2 Mm, the approximate location of the coronal base.

Although they were unable to extend their ER measurements below fluxes of ∼$4 \times 10^{18}$ Mx, Hagenaar et al. (2003) showed that the number of emerging ERs as a function of $\Phi_{ER}$ could be fitted with an exponential of the form $N_{ER}(\Phi_{ER}) \propto \exp[-\Phi_{ER}/(5 \times 10^{18} \text{ Mx})]$ (see their Figure 6). Scaling their minimum flux and e-folding constant upward by 1.7, it follows that

$$\frac{E_{ER}(\Phi_{ER} \gtrsim 1 \times 10^{18} \text{ Mx})}{E_{ER}(\Phi_{ER} \gtrsim 6.8 \times 10^{18} \text{ Mx})}$$
$$\simeq \frac{\int_{1 \times 10^{18} \text{ Mx}}^{\infty} \Phi_{ER} \exp[-\Phi_{ER}/(8.5 \times 10^{18} \text{ Mx})] d\Phi_{ER}}{\int_{6.8 \times 10^{18} \text{ Mx}}^{\infty} \Phi_{ER} \exp[-\Phi_{ER}/(8.5 \times 10^{18} \text{ Mx})] d\Phi_{ER}}$$
$$\simeq 1.23. \quad (4)$$

Combining Equations (2) and (4), we obtain as a lower bound on the flux emergence rate of ERs having $\Phi_{ER} \gtrsim 1 \times 10^{18}$ Mx:

$$E_{ER}(\Phi_{ER} \gtrsim 1 \times 10^{18} \text{ Mx}) \gtrsim 2.25 \times 10^{-3} \text{ Mx cm}^{-2} \text{ s}^{-1}. \quad (5)$$

As discussed below, the actual emergence rate may be higher because it depends on the time cadence of the magnetograph measurements.

## 3. Energy Flux Associated with ER Emergence

The ERs that emerge inside coronal holes will undergo interchange reconnection with the surrounding open flux, heating the corona and contributing to the acceleration of the solar wind (see, e.g., Parker 1991; Axford et al. 1999; Fisk et al. 1999; Cranmer & van Ballegooijen 2010; Cranmer 2018).

To estimate the contribution of ERs to coronal hole heating, let $B_0$ denote the average strength of the open flux at the coronal base. The timescale for an equal amount of minority-polarity flux to emerge into the corona is given by

$$\begin{aligned}
\tau_{recyc} &\sim \frac{B_0}{E_{ER}(\Phi_{ER} \gtrsim 1 \times 10^{18} \text{ Mx})/2} \\
&\sim 2.5 \left( \frac{B_0}{10 \text{ G}} \right) \left[ \frac{2.25 \times 10^{-3} \text{ Mx cm}^{-2} \text{ s}^{-1}}{E_{ER}(\Phi_{ER} \gtrsim 1 \times 10^{18} \text{ Mx})} \right] \text{hr},
\end{aligned} \quad (6)$$

where we note that only one half of the ER flux has polarity opposite to that of the open flux. Over this "recycling" or flux replacement time, all of the open flux will have undergone interchange reconnection, with magnetic energy being dissipated in a layer whose thickness is determined by the heights of the ER loops. Denoting the average loop height by $\langle h_{ER} \rangle$, we obtain for the energy flux density arising from interchange reconnection:

$$\begin{aligned}
F_{ER0} &\sim \frac{B_0^2}{8\pi} \left( \frac{\langle h_{ER} \rangle}{\tau_{recyc}} \right) \sim 2.9 \times 10^5 \left( \frac{B_0}{10 \text{ G}} \right) \\
&\times \left( \frac{\langle h_{ER} \rangle}{6.4 \text{ Mm}} \right) \left[ \frac{E_{ER}(\Phi_{ER} \gtrsim 1 \times 10^{18} \text{ Mx})}{2.25 \times 10^{-3} \text{ Mx cm}^2 \text{ s}^{-1}} \right] \text{erg cm}^{-2} \text{ s}^{-1}.
\end{aligned} \quad (7)$$

Here, the average loop height has been normalized to the value corresponding to $\Phi_{ER} = 1 \times 10^{19}$ Mx in Equation (3).

As indicated by Figure 6 of Hagenaar et al. (2008), the measured rate of ER emergence and thus the recycling time are very sensitive to the cadence of the magnetograms employed: as the time interval between successive magnetograms decreased from ∼1 hr to ∼5 minutes, $\tau_{recyc}$ decreased almost linearly from ∼13 to ∼1.5 hr. This suggests that Hagenaar et al. would have obtained higher emergence rates by using magnetograms taken at intervals less than $\Delta t = 5$ minutes, the cadence on which their 2008 study was based. Thus $F_{ER0}$ may be substantially higher than the estimate of ∼$2.9 \times 10^5$ erg cm$^{-2}$ s$^{-1}$ given by Equation (7).

## 4. Comparison with the Observed Solar Wind Energy Flux

The predicted ER energy flux and its scaling $F_{ER0} \propto B_0$ may be compared with observational estimates of the solar wind energy flux at the coronal base (see also Wang 2016b). Let $v$ denote the wind speed, $n_p$ the proton density, $\rho \simeq n_p m_p$ the mass density, and $B$ the strength of the radial field component. From the conservation of mass along a flux tube, the mass flux density at the coronal base (subscript "0") is related to that at Earth (subscript "E") by

$$\rho_0 v_0 = \left(\frac{B_0}{B_E}\right)\rho_E v_E. \quad (8)$$

Denoting by $F_w$ the nongravitational contribution to the total energy flux density, we also have from energy conservation:

$$\frac{F_{w0} - \rho_0 v_0 (GM_\odot/R_\odot)}{B_0} \simeq \frac{F_{wE}}{B_E} \simeq \frac{\rho_E v_E^3/2}{B_E}, \quad (9)$$

where we have neglected radiative losses and used the fact that the bulk kinetic energy provides the dominant contribution to $F_w$ at 1 au. The (nongravitational) energy flux density at the coronal base may then be written as

$$\begin{aligned}F_{w0} &\simeq \rho_0 v_0 \left(\frac{GM_\odot}{R_\odot} + \frac{1}{2}v_E^2\right) \\ &\simeq B_0 \left(\frac{\rho_E v_E}{B_E}\right)\left(\frac{GM_\odot}{R_\odot} + \frac{1}{2}v_E^2\right).\end{aligned} \quad (10)$$

Here, $\rho_E$, $v_E$, and $B_E$ are known from in situ measurements, while $B_0$ may be derived by applying a potential-field source-surface (PFSS) extrapolation to solar magnetograph data.

Let $r$ denote heliocentric distance, $L$ heliographic latitude, and $\phi$ Carrington longitude. In our version of the PFSS model, the magnetic field **B** remains current-free from near the photosphere to $r = R_{ss} = 2.5\,R_\odot$, where it becomes radial. At the inner boundary, $B_r$ is matched to the photospheric field, as given by 27.3 day synoptic maps from MWO and the Wilcox Solar Observatory (WSO), which have a longitudinal resolution of ~5° (~2 supergranular diameters). The MWO and WSO line-of-sight measurements are deprojected by dividing by $\cos L$, corrected for the saturation of the Fe I 525.0 nm line profile by multiplying by $(4.5 - 2.5\,\sin^2 L$; see Wang & Sheeley 1995), and averaged together. Tracing along field lines from $r = R_{ss}$ to $r = R_\odot$, we obtain a value of $B_0$ corresponding to every 5° of source-surface longitude. Because of the low spatial resolution of the magnetograph measurements, the field strength at the coronal base may be taken to be the same (by flux conservation) as that at the photospheric footpoint.

From hourly values extracted from the OMNIWeb database,[1] we calculate 8 hr averages of the near-Earth wind speed, proton density, and radial interplanetary field strength. The in situ data are mapped back to the source surface, taking into account the longitude shift due to solar rotation during the wind transit time ($\propto v_E^{-1}$), and then interpolated to match the 5° longitude spacing of the PFSS-derived $B_0$ values. Finally, Equations (8) and (10) are used to infer the mass and energy flux densities at the coronal base.

Figure 2 displays log–log scatter plots of (a) the proton flux density at the coronal base, $n_0 v_0$, against $B_0$, and (b) the energy flux density $F_{w0}$ against $B_0$, for the period 1998–2011. A data point is plotted for every 5° of longitude during Carrington rotations (CRs) 1933–2112. The dashed lines represent least-squares fits to the data, while the dotted lines indicate a slope of unity on the log–log scale. The mass and energy flux densities at the coronal base both increase almost linearly with the field strength, over a range of two orders of magnitude in $B_0$, $n_0 v_0$, and $F_{w0}$. The magnitude of $F_{w0}$ and its approximately linear scaling with $B_0$ are consistent with the rough estimate given by Equation (7) for $F_{ER0}$, the energy flux density associated with interchange reconnection between ERs and open field lines. Moreover, the scaling $n_0 v_0 \propto B_0$ is consistent with the idea that the solar wind mass flux is driven by the dissipation of magnetic energy near the coronal base.

---

[1] http://omniweb.gsfc.nasa.gov

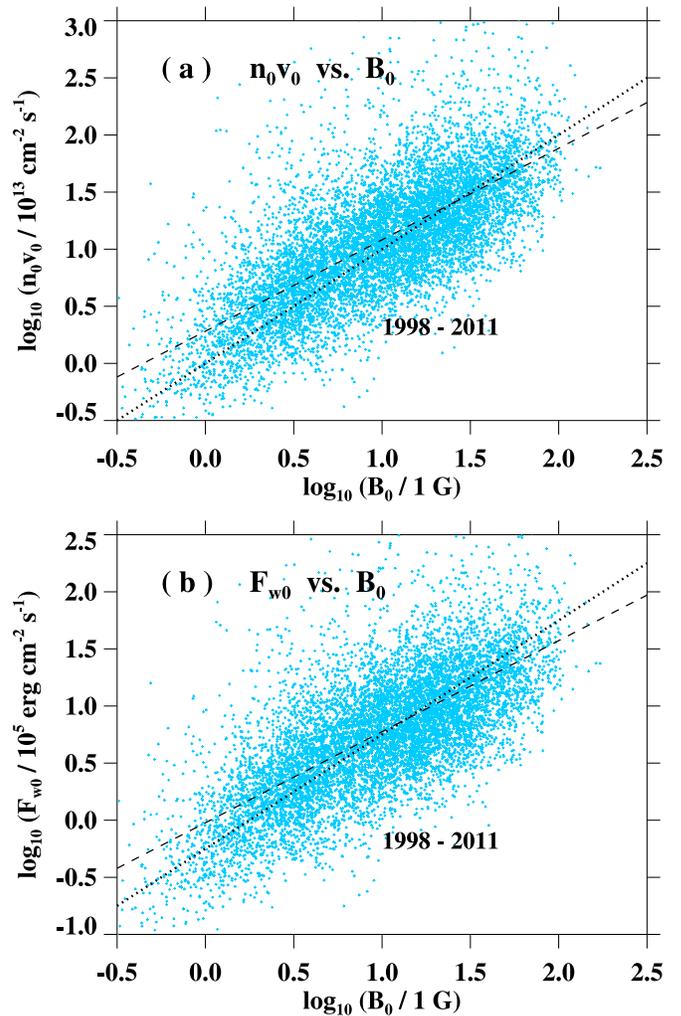

**Figure 2.** Scatter plots of (a) proton flux density at the coronal base, $\log_{10}(n_0 v_0)$ ($10^{13}$ cm$^{-2}$ s$^{-1}$), vs. base field strength, $\log_{10}(B_0)$ (G); and (b) (nongravitational) energy flux density at the coronal base, $\log_{10}(F_{w0})$ ($10^5$ erg cm$^{-2}$ s$^{-1}$), vs. $\log_{10}(B_0)$. A point is plotted for every 5° of source-surface longitude during CRs 1933–2112 (1998 February to 2011 July). In each log–log plot, the dashed line shows a least-squares fit to the data, while the dotted line indicates a slope of unity. The coronal base parameters were derived by applying a PFSS extrapolation with $R_{ss} = 2.5\,R_\odot$ to an average of MWO and WSO photospheric field maps, and were related to OMNIWeb measurements of $v_E$, $n_E$, and $B_E$ by using mass and energy conservation along a flux tube and taking into account the Sun–Earth transit time ($\propto v_E^{-1}$). Data from the Advanced Composition Explorer were employed to exclude intervals when the O$^{7+}$/O$^{6+}$ ratio exceeds 0.5, which are likely to be associated with coronal mass ejections.

## 5. Diffuse Plume Emission and the Origin of Low-frequency Alfvén Waves

Most of the MHD waves generated by convective motions at the photosphere are reflected as a result of the steep density gradients in the transition region, with only a small fraction of the wave energy leaking into the corona (see, e.g., Cranmer &

van Ballegooijen 2005). However, reconnection events that occur in the low corona will provide an additional source of Alfvén and magnetosonic waves that may contribute to the heating and acceleration of the solar wind.

Alfvén wave energy may be dissipated through nonlinear interactions between outgoing and reflected waves (Heinemann & Olbert 1980). Neglecting nonradial gradients in the Alfvén speed $v_A = B/(4\pi\rho)^{1/2}$, the condition for reflection in the sub-Alfvénic wind region ($r \lesssim 10\,R_\odot$) is that

$$P_{\rm wave} > P_{\rm crit} \simeq \frac{2\pi}{|\partial v_A/\partial r|}, \qquad (11)$$

where $P_{\rm wave}$ is the wave period. With $|\partial v_A/\partial r| \sim v_A/r$, $v_A \sim 1000$ km s$^{-1}$, and $r \sim R_\odot$, $P_{\rm crit}$ is on the order of an hour in the lower corona.

Coronal reconnection events may be driven by the emergence of ERs below them, or by footpoint motions due to granular or supergranular convection. ER emergence occurs on timescales on the order of 1 hr, and thus may generate waves that marginally satisfy the reflection criterion. Granular motions might be expected to give rise to Alfvén waves with characteristic periods on the order of the ~10 minute granule lifetime. Such relatively high-frequency waves could be reflected if they encountered large density inhomogeneities, such as those associated with slow-mode or compressional waves (Suzuki & Inutsuka 2006).

Reconnection on supergranular timescales ($\lesssim 1$ day) is associated with coronal/polar plumes. Pre-SDO observations suggested that plumes occupy only a small fraction of the coronal hole area (~10% according to Ahmad & Withbroe 1977). However, with the greatly improved sensitivity of the AIA instrument, it is now apparent that diffuse Fe IX 17.1 nm emission is present almost everywhere inside coronal holes. The brightest structures correspond to the "traditional" plumes identified in images from earlier instruments, but "plume haze" overlies every network concentration (see Figure 3).

As described in Wang et al. (2016; see also Avallone et al. 2018; Qi et al. 2019), plume emission brightens as supergranular flows converge, bringing network, intranetwork, and ER fluxes together to form dense clumps; as the underlying flows diverge again and the clumps are dispersed, the Fe IX emission gradually fades. During the convergence phase, which lasts several hours, the entrained minority-polarity flux is brought into increasingly close contact with the open flux that threads the dominant-polarity network elements, driving interchange reconnection at progressively faster rates. Individual bursts of reconnection may occur on the granular scale, giving rise to "jetlets" with quasi-periodicities of ~5–15 minutes (Tian et al. 2011; Raouafi & Stenborg 2014; Sheeley et al. 2014; Panesar et al. 2018). The bulk of the diffuse 17.1 nm emission, having a characteristic temperature of $T \sim 0.7$–0.8 MK, originates not from the jetlets themselves, but from the material evaporated from below as energy released by the ensemble of small-scale reconnection events is conducted downward through the transition region.

This quasi-continuous, cyclic driving of the interchange reconnection by the converging and diverging supergranular flows, as reflected in the gradually evolving Fe IX haze, may be a major source of Alfvén waves with periods ranging from hours to a day. Launching such waves from above the coronal

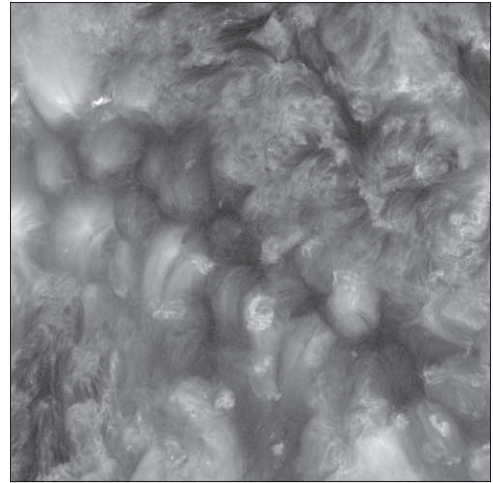
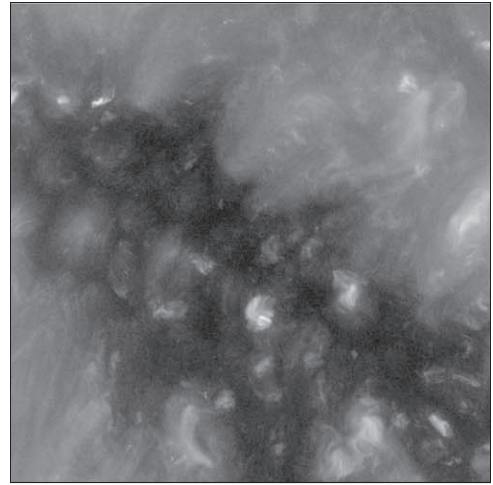
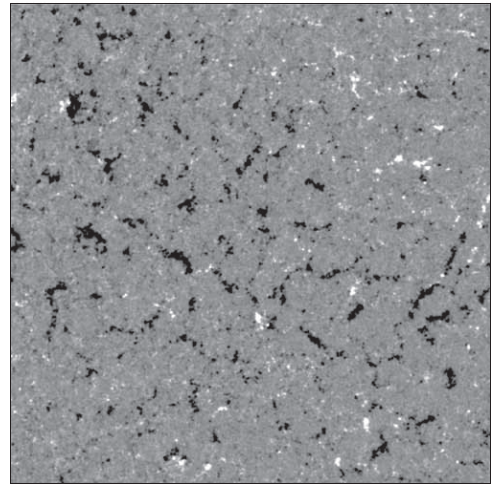

**Figure 3.** Equatorial coronal hole observed by SDO on 2012 June 29. Top panel: Fe IX 17.1 nm image taken at 05:00:36 UTC. Middle panel: Fe XIV 21.1 nm image taken at 05:00:36 UTC. Bottom panel: HMI line-of-sight magnetogram recorded at 05:00:41 UTC and saturated at ±50 G. Diffuse 17.1 nm emission is visible above all of the supergranular network concentrations inside the coronal hole, not just in "classical" coronal plumes such as the relatively bright structure at the upper left corner of the top frame. This suggests that supergranular flows are everywhere, continually driving reconnection between unipolar network elements and ER and intranetwork flux.

base allows them to circumvent the reflecting barrier represented by the transition region.

## 6. Alfvén Wave Reflection in the Corona

The amount of reflection and dissipation undergone by Alfvén waves depends on their periods and on the local gradients in the Alfvén speed (Equation (11)). It is thus important to know how $v_A$ varies along open flux tubes, which in turn requires a knowledge of the magnetic field and density distributions inside coronal holes. We now construct profiles of $v_A(r)$ and examine the behavior of $P_{crit}(r)$.

Although the PFSS model provides a reasonable approximation to the coronal field in the low-$\beta$ regime below the source surface, it fails to take into account the heliospheric sheet currents located beyond $r \sim 2.5\ R_\odot$, which act to isotropize the flux distribution. We therefore employ a more realistic "potential-field current-sheet" model, in which the field is current-free everywhere except at a current sheet whose inner edge is at $r = 2.5\ R_\odot$. For simplicity, the photospheric flux distribution is assigned the axisymmetric form

$$B_r(R_\odot, L) = B_0 \sin^q L, \qquad (12)$$

where $q$ is a positive odd integer. Laplace's equation is then solved subject to this inner boundary condition and to the requirement that $B_L(r, L = 0°) = 0$ for $r > 2.5\ R_\odot$ (see Sheeley et al. 1989). As illustrative examples, we calculate the variation of the field along the polar axis $L = 90°$ and along a flux tube that is directed toward the equatorial plane and passes through the point $(r = 215 R_\odot, L = 89°)$. In both cases, we set $B_0 = 10$ G and $q = 7$, so that the photospheric field has a topknot form similar to that observed near solar minimum and the area of open flux extends from the pole to latitude 62°, as in a typical polar coronal hole.

For the radial variation of the density, $n(r)$, we consider a variety of forms. Along the polar axis, we employ the "plume" and "interplume" densities plotted in Figure 2 of Wang (1994a); these model solar wind density profiles were fitted to the Mg X plume measurements of Ahmad & Withbroe (1977) and to the EUV observations of a polar coronal hole by Mariska (1978), respectively. Along the boundary field line, we use the "model S1" slow-wind densities plotted in Figure 3 of Wang (1994b). Alternatively, for both the polar and boundary field lines, we adopt a slightly modified version of the empirical density falloff given by Cranmer & van Ballegooijen (2005), which was based partly on white-light polarization measurements ("CvB" model):

$$n(\hat{r}) = 6.45 \times 10^7 \left( \frac{1}{\hat{r}^{32}} + \frac{3}{\hat{r}^{16}} + \frac{0.6}{\hat{r}^8} + \frac{0.05}{\hat{r}^4} + \frac{0.002}{\hat{r}^2} \right) \mathrm{cm}^{-3}. \qquad (13)$$

Here $\hat{r} \equiv r/R_\odot$ and we have adjusted the leading term so that $n = 3 \times 10^8\ \mathrm{cm}^{-3}$ at the coronal base $\hat{r} = 1$.

With the photospheric field set to $B_r(R_\odot, L) = 10\ \mathrm{G} \sin^7 L$ and the densities given by the plume, interplume, and CvB models, Figure 4 shows the variation of $v_A(r)$ and $n(r)$ along the polar-axis field line. Also plotted is

$$f_{exp}(r) = \left(\frac{R_\odot}{r}\right)^2 \left| \frac{B_0}{B_r(r)} \right|, \qquad (14)$$

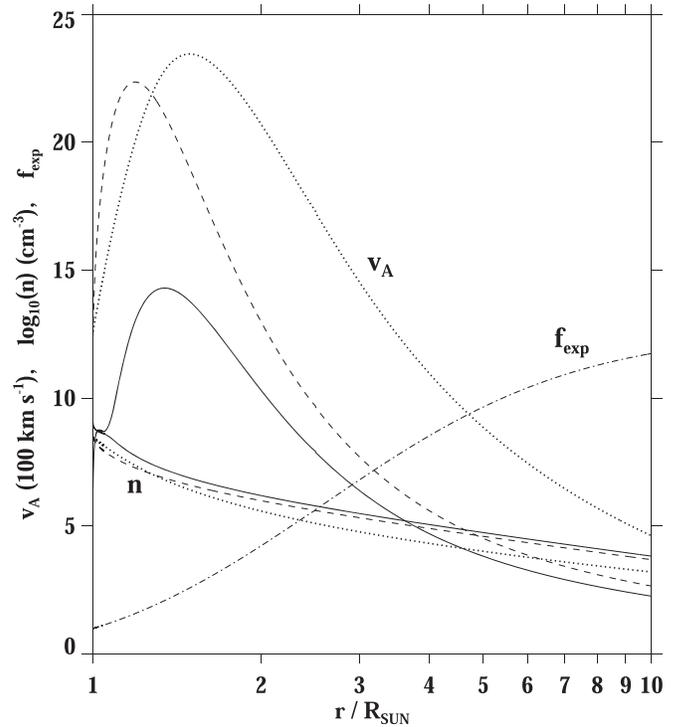

**Figure 4.** Radial variation of the Alfvén speed $v_A = B/(4\pi\rho)^{1/2}$ along a polar-axis field line. The photospheric flux distribution has the axisymmetric topknot form $B_r(R_\odot, L) = 10\ \mathrm{G} \sin^7 L$, resembling that typically observed near solar minimum; the coronal field is current-free everywhere except at an equatorial current sheet whose inner edge is located at $r = 2.5\ R_\odot$. The three curves correspond to different choices of the density distribution $n(r)$ (also plotted). Solid lines: densities from plume model PL3 of Wang (1994a). Dashed lines: densities from the interplume model IPL of Wang (1994a). Dotted lines: densities based on white-light observations (CvB model) and given by Equation (13). The dashed–dotted line shows the variation of the flux-tube expansion factor, $f_{exp}(r) = (R_\odot/r)^2 |B_0/B_r|$. Because $n^{1/2}$ initially falls off more steeply than $B$, $v_A(r)$ attains a maximum at $r \sim 1.2$–1.5 $R_\odot$.

the factor by which the flux tube expands in solid angle between the coronal base and heliocentric distance $r$. Because $n^{1/2}(r)$ initially falls off more rapidly than $B(r)$, $v_A(r) \propto B/n^{1/2}$ rises to a peak in the region $r \sim 1.2$–1.5 $R_\odot$ before declining again. Note also that $f_{exp}(r)$ continues to increase and $B(r)$ to fall off more steeply than $r^{-2}$ beyond $r = 2.5\ R_\odot$, unlike in the PFSS model.

Figure 5 displays, for each of the density models, the radial variation of the critical period for Alfvén wave reflection, $P_{crit} \propto |dv_A/dr|^{-1}$. Corresponding to the maxima in $v_A(r)$ seen in Figure 4, $P_{crit}(r)$ exhibits singularities in the region $r \sim 1.2$–1.5 $R_\odot$. Near the coronal base, there are additional singularities associated with the narrow density spike produced by the extra base heating in the plume model (see Wang 1994a). As $v_A$ decreases from its peak values, $P_{crit}$ increases from $\sim$1–2 hr at $r \sim 1.2$–2.4 $R_\odot$ to $\sim$5–10 hr at $r \sim 3$–6 $R_\odot$, beyond which it continues to increase monotonically with distance. In the inner corona where $dv_A/dr > 0$, Alfvén waves with periods significantly less than an hour may be reflected. Since $P_{crit} \propto B_0^{-1}$, scaling the footpoint field strength upward (downward) by a given factor will shift the curves downward (upward) by the same factor.

Figure 6 shows the variation of the Alfvén speed along the boundary field line originating at latitude 62°.3, along with the adopted density profiles from model "S1" of Wang (1994b) and from Equation (13). The expansion factor $f_{\rm exp}(r)$ (also plotted) has a sharp peak at $r \sim 2.5~R_\odot$, where the flux tube approaches the null point at the inner edge of the equatorial current sheet. Correspondingly, $v_{\rm A}(r)$ falls to a minimum at this location, subsequently rising to a secondary maximum at $r \sim 4.75~R_\odot$ as $B(r)$ increases and then starts to decrease again. As in the case of the polar flux tube, $v_{\rm A}(r)$ also attains a maximum in the lower corona as a result of the steep initial falloff of $n(r)$.

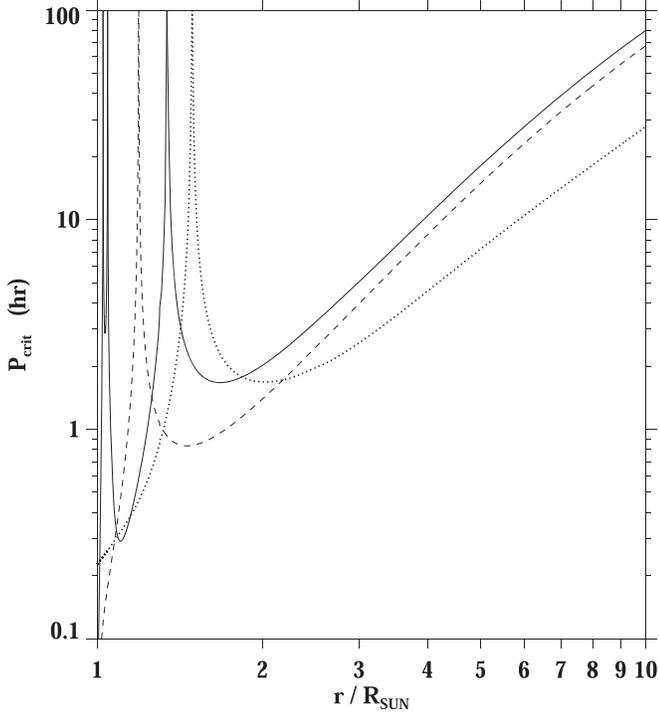

**Figure 5.** Radial variation of the critical period for Alfvén wave reflection, $P_{\rm crit} \simeq 2\pi/|dv_{\rm A}/dr|$, along the polar-axis field line. The photospheric flux distribution again has the form $B_r(R_\odot, L) = 10~{\rm G}\sin^7 L$. Solid line: densities from plume model PL3. Dashed line: densities from interplume model IPL. Dotted line: densities given by Equation (13). The singularities in the region $r \sim 1.2$–$1.5~R_\odot$ correspond to the maxima in $v_{\rm A}(r)$ seen in Figure 4. The critical period for the plume model shows additional singularities associated with the density spike due to enhanced heating at the coronal base.

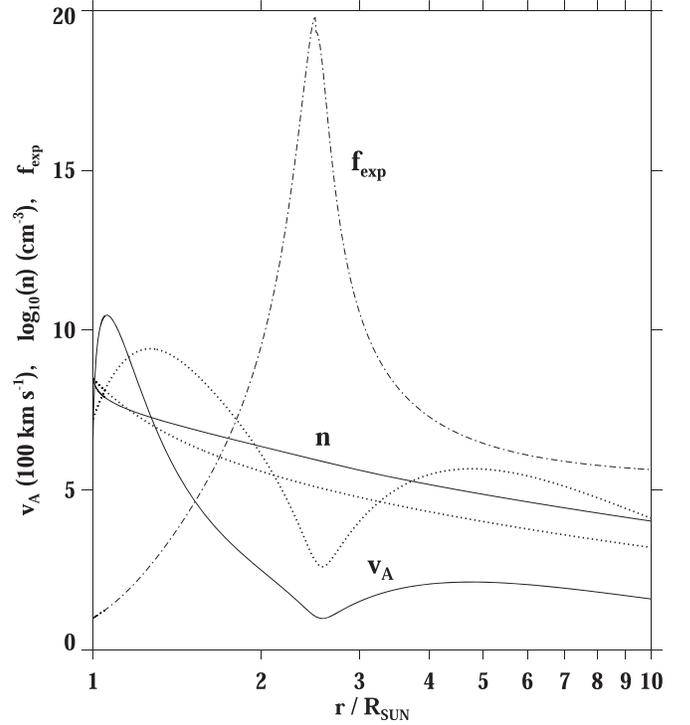

**Figure 6.** Radial variation of the Alfvén speed $v_{\rm A}(r)$ and density $n(r)$ along a "boundary" flux tube originating at latitude $62°.3$. The photospheric flux distribution again has the form $B_r(R_\odot, L) = 10~{\rm G}\sin^7 L$, and the coronal field is potential everywhere except at an equatorial current sheet that extends outward from $r = 2.5~R_\odot$. Solid curves: densities from the slow-wind model S1 of Wang (1994b). Dotted curves: densities given by the CvB model (Equation (13)). Also plotted is the expansion factor $f_{\rm exp}(r)$ (dashed–dotted line), which rises to a sharp peak near the null point at the inner edge of the current sheet. As the flux tube traverses this region, $v_{\rm A}(r)$ goes through a local minimum; it then rises to a local maximum at $r \sim 4.75~R_\odot$, before continuing to decrease monotonically. As in the case of the monotonically expanding polar flux tube, the main maximum in $v_{\rm A}(r)$ occurs in the lower corona.

Corresponding to the three extrema in $v_{\rm A}(r)$, $P_{\rm crit}(r) \propto |\partial v_{\rm A}/\partial r|^{-1}$ shows singularities at $r \sim 1.05$–$1.25~R_\odot$, $r \sim 2.5~R_\odot$, and $r \sim 4.75~R_\odot$ (Figure 7). Beyond $r \sim 4~R_\odot$, only waves with periods longer than a day are reflected.

## 7. Flux-tube Expansion and the Radial Dependence of Coronal Hole Heating

Interchange reconnection driven by the emergence and cancellation of ER flux will give rise to localized ohmic heating in the low corona. In addition, heat conduction, plasma jets, and MHD waves will transport energy inward and outward from the reconnection sites. The asymptotic wind speed will depend on the amount of energy that is deposited in the outer corona (near and beyond the sonic point) relative to that dissipated near the coronal base (Leer & Holzer 1980). We now discuss how the rate of flux-tube expansion affects the radial distribution of the heating, and thus the wind speed.

### 7.1. Flux-tube Expansion and Heat Conduction

If radiative losses are neglected, the conservation of energy along a flux tube may be written schematically as

$$\frac{F_{\rm c} + F_{\rm m} + F_{\rm A}}{B} + \zeta\left(\frac{5kT}{m_{\rm p}} + \frac{1}{2}v^2 - \frac{GM_\odot}{r}\right) \simeq {\rm constant}, \quad (15)$$

where $F_{\rm c}$, $F_{\rm m}$, and $F_{\rm A}$ represent the conductive, mechanical/nonwave, and Alfvén/MHD-wave energy flux densities, respectively, $T$ denotes the electron temperature, and $\zeta \equiv \rho v/B$ is constant along the flux tube. In the classical, collision-dominated approximation,

$$F_{\rm c} \simeq -\kappa T^{5/2}\frac{dT}{ds}, \quad (16)$$

where $\kappa$ is the heat conductivity and $s$ is the length coordinate along $B$. For simplicity, we henceforth consider a radially oriented flux tube along which

$$B(r) = B_0\left(\frac{R_\odot}{r}\right)^\nu, \quad (17)$$

where $\nu \geqslant 2$.

Above the reconnection region (within which most of the "mechanical" energy is deposited), the conductive energy flux will be directed outward, with the net flux (integrated over the areal cross section of the flux tube) scaling as

$$\frac{F_c}{B} \propto -r^\nu T^{5/2} \frac{dT}{dr} \sim r^{\nu-1} T^{7/2}. \quad (18)$$

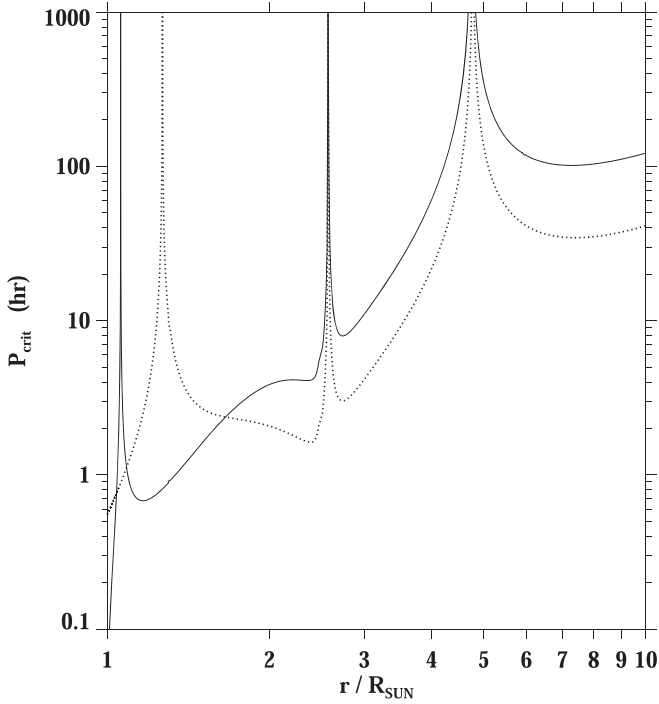

**Figure 7.** Radial variation of the critical period for Alfvén wave reflection, $P_{\rm crit}(r)$, along the boundary flux tube that skirts the inner edge of the equatorial current sheet. Solid curve: densities from slow-wind model S1. Dotted curve: densities given by Equation (13). Corresponding to the two peaks in $v_A(r)$ and the intervening dip near the null point (Figure 6), $P_{\rm crit}$ exhibits singularities at $r \sim 1.05$–$1.25\ R_\odot$, $r \sim 2.5\ R_\odot$, and $r \sim 4.75\ R_\odot$.

If we assume that the rate of energy dissipation in the reconnection region is independent of $\nu$, then Equation (18) indicates that $T$ must decrease as $\nu$ increases, in order for the net conductive flux to remain the same. If the wind is mainly thermally driven, this in turn implies that the wind speed will decrease as the expansion factor increases. These arguments are supported by Figure 5 of Grappin et al. (2011), which compares plume solutions for which $\nu = 2$ and $\nu = 5$. In these two-temperature simulations, the temperatures and velocities are much lower beyond $r \sim 1.5\ R_\odot$ along the more rapidly expanding flux tube, even though identical "base" and "extended" heating functions were assigned in both cases.

The effect of varying $\nu$ on the conductive flux and the wind speed is consistent with the observed inverse correlation between $v_E$ and the expansion factor at the source surface,

$$f_{\rm ss} \equiv f_{\rm exp}(R_{\rm ss}) = \left(\frac{R_\odot}{R_{\rm ss}}\right)^2 \frac{B_0}{B_{\rm ss}}, \quad (19)$$

where $B_{\rm ss} \equiv |B_r(R_{\rm ss})|$ (Wang & Sheeley 1990; Arge & Pizzo 2000; Fujiki et al. 2015; Poduval 2016; Reiss et al. 2019, 2020; see also Figure 8(c)).

### 7.2. Flux-tube Expansion and Turbulent Wave Heating

Alfvén waves with periods $P_{\rm wave} > P_{\rm crit}(r)$ are reflected and undergo nonlinear interaction with outgoing waves, resulting in the turbulent dissipation of energy. A widely adopted phenomenological form for the damping rate is

$$Q_{\rm turb} \simeq \rho \left( \frac{Z_-^2 Z_+ + Z_+^2 Z_-}{L_\perp} \right) \quad (20)$$

(see, e.g., Dmitruk et al. 2001, 2002; Cranmer & van Ballegooijen 2005; Cranmer 2010). In this expression, $L_\perp$ denotes the perpendicular correlation length for the largest turbulent eddies, while $Z_-$ and $Z_+$ represent the frequency-averaged, rms values of the Elsässer variables:

$$Z_\pm \equiv \langle \delta v \pm \delta B / (4\pi\rho)^{1/2} \rangle, \quad (21)$$

where $\delta v$ and $\delta B$ are the amplitudes of the transverse velocity and magnetic oscillations. Since the incoming wave amplitude $Z_+$ is expected to be small compared with the outgoing amplitude $Z_-$, the volumetric heating rate may be approximated as

$$Q_{\rm turb} \simeq \rho \frac{Z_-^2 Z_+}{L_\perp}. \quad (22)$$

As shown by Dmitruk et al. (2001, 2002), in the strong turbulence limit ($L_\perp \to 0$), $Z_+/L_\perp = |dv_A/dr|$ and the Alfvén wave energy flux density $F_A \simeq \rho v_A Z_-^2$ decays as

$$B \frac{d}{dr}\left( \frac{F_A}{B} \right) \simeq -Q_{\rm turb} \simeq -\rho Z_-^2 \left| \frac{dv_A}{dr} \right| \simeq -\frac{1}{v_A} \left| \frac{dv_A}{dr} \right| F_A. \quad (23)$$

Solving for $F_A/B$ then yields

$$\frac{F_A}{B} \propto \begin{cases} v_A^{-1}, & dv_A/dr > 0, \\ v_A, & dv_A/dr < 0, \end{cases} \quad (24)$$

and

$$Q_{\rm turb} \propto \begin{cases} \dfrac{B}{v_A^2} \left| \dfrac{dv_A}{dr} \right|, & dv_A/dr > 0, \\ B \left| \dfrac{dv_A}{dr} \right|, & dv_A/dr < 0. \end{cases} \quad (25)$$

Equation (24) indicates that the net (area-integrated) Alfvén wave energy flux is damped at a rate proportional to $v_A(r)$ in the region where $dv_A/dr < 0$. Thus, as the expansion factor increases and $v_A$ falls off more steeply, the amount of wave energy dissipated well inside the sonic point (located at $r \sim 2$–$6\ R_\odot$) will increase relative to that deposited near and beyond the sonic point. This is again consistent with the empirical inverse correlation between asymptotic wind speed and the rate of coronal flux-tube expansion.

### 7.3. Solar Wind Speed and the Source-surface Field Strength

In the model of Suzuki & Inutsuka (2006), the damping rate of Alfvén waves depends on the nonlinearity parameter $Z_-/v_A$:

$$Q_{\rm SI} \propto \frac{Z_-}{v_A} \equiv \frac{\langle \delta v - \delta B/(4\pi\rho)^{1/2} \rangle}{v_A}. \quad (26)$$

If $v_A$ falls off slowly along a given flux tube, $Z_-/v_A$ remains small and the waves are dissipated over longer distances,

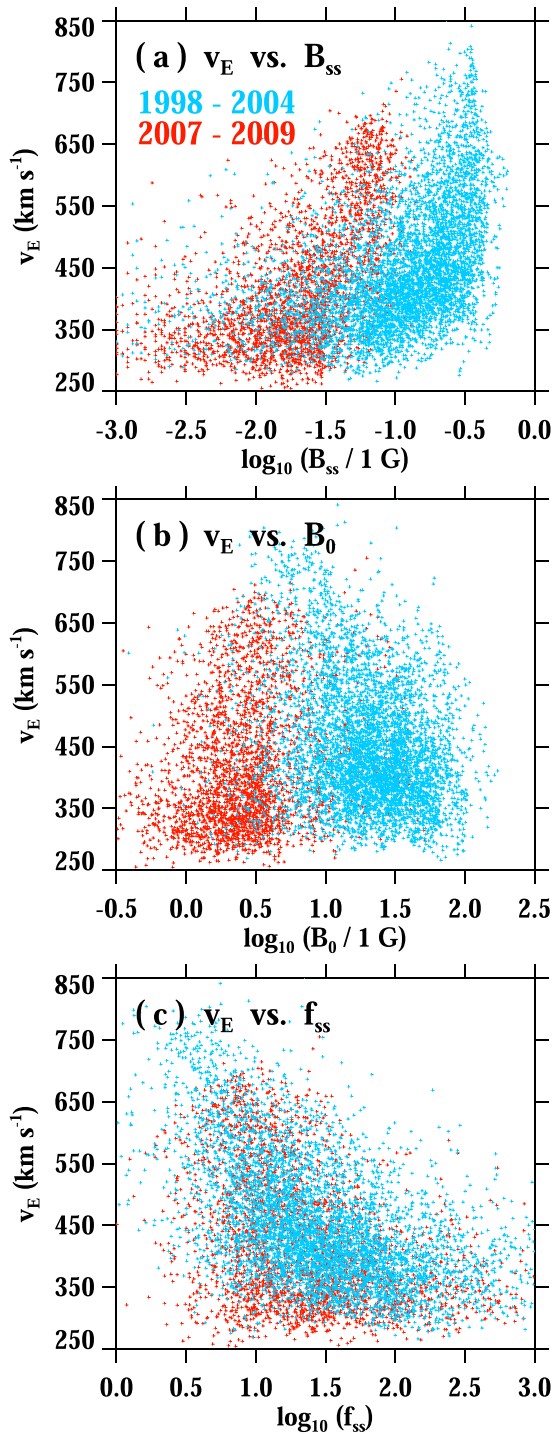

**Figure 8.** Scatter plots of near-Earth wind speed $v_E$ against (a) the field strength at the source surface $R_{ss} = 2.5\,R_\odot$, $\log_{10}(B_{ss})$ (G); (b) the field strength at the coronal base, $\log_{10}(B_0)$ (G); and (c) the expansion factor at the source surface, $\log_{10}(f_{ss})$. A point is plotted for every 5° of source-surface longitude, with blue dots representing the period 1998–2004 (CRs 1933–2024), and red dots representing 2007–2009 (CRs 2052–2091). Again, a PFSS extrapolation of MWO and WSO photospheric field maps was used to relate the in situ measurements of $v_E$ to the coronal field parameters, excluding intervals when $O^{7+}/O^{6+} > 0.5$. The source-surface field strengths corresponding to a given wind speed are seen to be systematically smaller near solar minimum than near solar maximum. In contrast, the inverse relationship between $v_E$ and the flux-tube expansion factor is independent of solar cycle phase. Contrary to the prediction of Fisk et al. (1999), $v_E$ is essentially uncorrelated with the field strength at the coronal base.

leading to more energy deposition near the sonic point and faster wind. In this model, $v_E$ is correlated with the parameter

$$\frac{B_0}{f_{ss}} \propto B_{ss}, \quad (27)$$

and thus scales as the field strength at the source surface.

As indicated by Figure 8, however, this prediction is not supported by observations. Here we have plotted the near-Earth wind speed against $B_{ss}$, $B_0$, and $f_{ss}$, using PFSS extrapolations of MWO/WSO photospheric field maps to trace along Earth-directed field lines, as in Section 3; the source surface is again located at $r = R_{ss} = 2.5\,R_\odot$. Blue (red) dots represent data from 1998–2004 (2007–2009), with a point being plotted for every 5° of source-surface longitude during CRs 1933–2024 (2052–2091). Although the scatter plot of $v_E$ versus $B_{ss}$ (Figure 8(a)) shows a clear tendency for the wind speed to increase with the source-surface field strength, the points for the sunspot minimum period are shifted toward systematically weaker fields than the points for 1998–2004. Thus $v_E$ cannot be a function of $B_{ss}$ alone (see also Fujiki et al. 2015).

The scatter plot of $v_E$ against the field strength $B_0$ at the coronal base (Figure 8(b)) shows almost no correlation between these quantities during either period, contrary to the prediction of Fisk et al. (1999). In contrast, the tendency for $v_E$ to decrease with increasing $f_{ss}$ is seen to hold independently of solar cycle phase, with no systematic separation between the blue and red dots (Figure 8(c)).

Although the turbulent dissipation rate given by Equation (20) appears to be consistent with the empirical inverse correlation between $v_E$ and $f_{ss}$, the validity of this formula has been questioned by Roberts (2010) on both observational and theoretical grounds. In addition, van Ballegooijen & Asgari-Targhi (2016) and Verdini et al. (2019) found that it significantly overestimates the dissipation associated with incompressible turbulence, and that compressive effects must be introduced to boost the heating rate. However, the $v_E$–$f_{ss}$ relationship may be explained by a wide variety of heating mechanisms, provided that a steeper falloff in $B(r)$ causes the rate of energy dissipation per unit mass to decrease more rapidly with heliocentric distance. Equivalently, we require that

$$\frac{F_{\text{heating}}}{B} \propto B^\alpha(r), \quad (28)$$

where $\alpha > 0$ and $dB/dr < 0$.

## 8. Summary and Discussion

Our conclusions may be summarized as follows:

1. The frequent observation of loop-like fine structure in EUV images without corresponding minority-polarity signatures in magnetograms suggests that the rate of ER emergence inside unipolar regions has been substantially underestimated.

2. On the assumption that ERs emerge at the same rate inside coronal holes as in the quiet Sun, the energy flux density associated with interchange reconnection is found to be on the order of $10^5\,\text{erg cm}^{-2}\,\text{s}^{-1}$ and increase almost linearly with $B_0$, the field strength at the coronal base. The predicted behavior is consistent with the energy flux densities inferred from in situ solar wind observations and PFSS extrapolations.

3. Reconnection between the ER loops and the open flux is driven by the emergence process itself (on timescales on the order of 1 hr), by granular convection (on timescales of ∼5–15 minutes), and by supergranular flows (on timescales of several hours to a day). The multiple small-scale reconnection events give rise to ohmic heating, jets/jetlets, and MHD waves. Inward heat conduction from the reconnection sites leads to chromospheric evaporation and helps to drive the solar wind mass flux.

4. The presence of diffuse Fe IX 17.1 nm emission everywhere above the magnetic network inside coronal holes is taken as evidence for reconnection on supergranular timescales, which in turn may be a major source of Alfvén waves with periods of several hours to a day.

5. Models for the variation of the Alfvén speed near the Sun indicate that the critical period for wave reflection, $P_{\rm crit} \simeq 2\pi/|\partial v_{\rm A}/\partial r|$, increases from ∼1–2 hr at $r \sim 1.1$–2.4 $R_\odot$ to over a day for $r \gtrsim$ 4–6 $R_\odot$. Shorter-period Alfvén waves may be reflected and dissipated in the inner corona ($r \lesssim 1.05$–1.3 $R_\odot$), where $\partial v_{\rm A}/\partial r > 0$.

6. Heat is conducted outward as well as inward from the temperature maximum in the reconnection region. Increasing the rate of flux-tube expansion has the effect of decreasing the outward conductive flux density and thus the coronal temperature, resulting in lower asymptotic wind speeds for a given rate of energy deposition near the coronal base.

7. Increasing the flux-tube expansion factor also causes the amount of turbulent energy dissipation of Alfvén waves to fall off more steeply with heliocentric distance, so that relatively more energy is deposited well inside the sonic point, again leading to lower asymptotic wind speeds.

8. Increasing the field strength at the coronal base acts to increase the local mass flux density (Figure 2(a)), but has little systematic effect on the final wind speed (Figure 8(b)).

In our "hybrid" model, interchange reconnection driven by ER emergence provides the main heating source for coronal holes, while Alfvén waves generated in the reconnection process propagate outward and help to accelerate the solar wind. As argued by Roberts (2010), Alfvénic fluctuations and their turbulent dissipation may not alone be sufficient to power the wind. Verdini et al. (2010) have also concluded that an additional compressive heating is required near the coronal base.

From their wavelet decomposition of a long time series of COR1 white-light images, Telloni et al. (2013) identified fluctuating, raylike structures with quasi-periods and lifetimes of several hours. They interpreted these structures either as signatures of magnetic reconnection on supergranular timescales, or as long-wavelength density oscillations transverse to the magnetic field and sky plane. Their observations provide support for the idea that supergranular flows acting on ER and intranetwork fields may give rise to persistent, quasi-continuous reconnection in the corona and/or to MHD waves with very low frequencies.

Using the Ultraviolet Coronagraph Spectrometer on the Solar and Heliospheric Observatory, Antonucci et al. (2005) showed that the O VI and H I kinetic temperatures were substantially lower just inside the boundaries of polar coronal holes than near their centers. Consistent with the present analysis, these authors attributed the temperature difference to the more rapid flux-tube expansion occurring near the edges of the holes.

As suggested by Cranmer & van Ballegooijen (2005), a small fraction of the Alfvén waves generated by granular motions at the photosphere may leak out into the corona, providing an additional energy source for the solar wind. Again, however, although Alfvén waves originating from either below or above the coronal base may well inject energy and momentum at greater distances, interchange reconnection and the accompanying ohmic dissipation provide a far more effective means of heating the lower corona and driving the mass flux (see Parker 1991).

We have shown how differences in the wind speed may be explained by variations in the flux-tube expansion factor, without requiring a noncoronal hole source for the bulk of the slow solar wind. This is supported not only by the scatter plot of Figure 8(c), but also by observations of low-speed wind with high Alfvénicity, both at 1 au (e.g., D'Amicis et al. 2019; Wang & Ko 2019) and closer to the Sun (e.g., Bale et al. 2019; Stansby et al. 2019, 2020). However, the analysis presented here does not apply to the slow wind in the immediate vicinity of the heliospheric current sheet, which originates from the underlying helmet streamers. Here, the material is released through reconnection among the streamer loops themselves to produce flux ropes or "blobs" (Sheeley et al. 2009), and by interchange reconnection between these large-scale loops and the adjacent open flux, giving rise to the heliospheric plasma sheet (Wang et al. 1998). In contrast, the interchange reconnection that occurs inside coronal holes involves much smaller loops with scale sizes of a few megameters.

The scenario advocated here is based on the assumption that the rate of ER emergence inside coronal holes has been underestimated by a factor of 3 or more. In the near future, this assumption may be tested using the unprecedented sensitivity and spatial resolution of the Daniel K. Inouye Solar Telescope's visible and near-infrared spectropolarimeters. Also of importance for understanding the relative roles of reconnection and wave heating are measurements of the period distribution of Alfvén waves, and its radial evolution near the Sun, by the Parker Solar Probe.

I am grateful to S. R. Cranmer, R. Grappin, and D. A. Roberts for helpful correspondence. This work was supported by NASA and the Office of Naval Research.

ORCID iDs

Y.-M. Wang  https://orcid.org/0000-0002-3527-5958

References

Abramenko, V. I., Fisk, L. A., & Yurchyshyn, V. B. 2006, ApJL, 641, L65
Ahmad, I. A., & Withbroe, G. L. 1977, SoPh, 53, 397
Antonucci, E., Abbo, L., & Dodero, M. A. 2005, A&A, 435, 699
Arge, C. N., & Pizzo, V. J. 2000, JGR, 105, 10465
Avallone, E. A., Tiwari, S. K., Panesar, N. K., Moore, R. L., & Winebarger, A. 2018, ApJ, 861, 111
Axford, W. I., McKenzie, J. F., Sukhorukova, G. V., et al. 1999, SSRv, 87, 25
Bale, S. D., Badman, S. T., Bonnell, J. W., et al. 2019, Natur, 576, 237
Belcher, J. W., & Davis, L., Jr. 1971, JGR, 76, 3534
Cranmer, S. R. 2010, ApJ, 710, 676
Cranmer, S. R. 2018, ApJ, 862, 6
Cranmer, S. R., & van Ballegooijen, A. A. 2005, ApJS, 156, 265
Cranmer, S. R., & van Ballegooijen, A. A. 2010, ApJ, 720, 824
Cranmer, S. R., van Ballegooijen, A. A., & Edgar, R. J. 2007, ApJS, 171, 520
D'Amicis, R., Matteini, L., & Bruno, R. 2019, MNRAS, 483, 4665
Dmitruk, P., Matthaeus, W. H., Milano, L. J., et al. 2002, ApJ, 575, 571
Dmitruk, P., Milano, L. J., & Matthaeus, W. H. 2001, ApJ, 548, 482

Fisk, L. A. 2005, ApJ, 626, 563
Fisk, L. A., Schwadron, N. A., & Zurbuchen, T. H. 1999, JGR, 104, 19765
Fujiki, K., Tokumaru, M., Iju, T., Hakamada, K., & Kojima, M. 2015, SoPh, 290, 2491
Grappin, R., Wang, Y.-M., & Pantellini, F. 2011, ApJ, 727, 30
Hagenaar, H. J. 2001, ApJ, 555, 448
Hagenaar, H. J., DeRosa, M. L., & Schrijver, C. J. 2008, ApJ, 678, 541
Hagenaar, H. J., DeRosa, M. L., & Schrijver, C. J. 2010, ApJ, 715, 696
Hagenaar, H. J., Schrijver, C. J., & Title, A. M. 2003, ApJ, 584, 1107
Heinemann, M., & Olbert, S. 1980, JGR, 85, 1311
Hollweg, J. V. 1986, JGR, 91, 4111
Hollweg, J. V., & Isenberg, P. A. 2007, JGRA, 112, A08102
Leer, E., & Holzer, T. E. 1980, JGR, 85, 4681
Linker, J. A., Caplan, R. M., Downs, C., et al. 2017, ApJ, 848, 70
Mariska, J. T. 1978, ApJ, 225, 252
Moore, R. L., Musielak, Z. E., Suess, S. T., & An, C.-H. 1991, ApJ, 378, 347
Panesar, N. K., Sterling, A. C., Moore, R. L., et al. 2018, ApJL, 868, L27
Paraschiv, A. R., Bemporad, A., & Sterling, A. C. 2015, A&A, 579, A96
Parker, E. N. 1991, ApJ, 372, 719
Poduval, B. 2016, ApJL, 827, L6
Qi, Y., Huang, Z., Xia, L., et al. 2019, SoPh, 294, 92
Raouafi, N.-E., & Stenborg, G. 2014, ApJ, 787, 118
Reiss, M. A., MacNeice, P. J., Mays, L. M., et al. 2019, ApJS, 240, 35
Reiss, M. A., MacNeice, P. J., Muglach, K., et al. 2020, ApJ, 891, 165
Roberts, D. A. 2010, ApJ, 711, 1044
Sako, N., Shimojo, M., Watanabe, T., & Sekii, T. 2013, ApJ, 775, 22
Sheeley, N. R., Jr., Lee, D. D.-H., Casto, K. P., Wang, Y.-M., & Rich, N. B. 2009, ApJ, 694, 1471
Sheeley, N. R., Jr., Wang, Y.-M., & Harvey, J. W. 1989, SoPh, 119, 323
Sheeley, N. R., Jr., Warren, H. P., Lee, J., et al. 2014, ApJ, 797, 131
Stansby, D., Horbury, T. S., & Matteini, L. 2019, MNRAS, 482, 1706
Stansby, D., Matteini, L., Horbury, T. S., et al. 2020, MNRAS, 492, 39
Suzuki, T. K., & Inutsuka, S. 2006, JGRA, 111, A06101
Telloni, D., Ventura, R., Romano, P., Spadaro, D., & Antonucci, E. 2013, ApJ, 767, 138
Thornton, L. M., & Parnell, C. E. 2011, SoPh, 269, 13
Tian, H., McIntosh, S. W., Habbal, S. R., & He, J. 2011, ApJ, 736, 130
Tran, T., Bertello, L., Ulrich, R. K., & Evans, S. 2005, ApJS, 156, 295
Ulrich, R. K., Evans, S., Boyden, J. E., & Webster, L. 2002, ApJS, 139, 259
van Ballegooijen, A. A., & Asgari-Targhi, M. 2016, ApJ, 821, 106
Velli, M. 1993, A&A, 270, 304
Verdini, A., Grappin, R., & Montagud-Camps, V. 2019, SoPh, 294, 65
Verdini, A., Velli, M., Matthaeus, W. H., Oughton, S., & Dmitruk, P. 2010, ApJL, 708, L116
Verdini, A., Velli, M., & Oughton, S. 2005, A&A, 444, 233
Wang, Y.-M. 1994a, ApJL, 435, L153
Wang, Y.-M. 1994b, ApJL, 437, L67
Wang, Y.-M. 2016a, ApJL, 820, L13
Wang, Y.-M. 2016b, ApJ, 833, 121
Wang, Y.-M., & Ko, Y.-K. 2019, ApJ, 880, 146
Wang, Y.-M., & Sheeley, N. R., Jr. 1990, ApJ, 355, 726
Wang, Y.-M., & Sheeley, N. R., Jr. 1995, ApJL, 447, L143
Wang, Y.-M., Sheeley, N. R., Jr., Walters, J. H., et al. 1998, ApJL, 498, L165
Wang, Y.-M., Ugarte-Urra, I., & Reep, J. W. 2019, ApJ, 885, 34
Wang, Y.-M., Warren, H. P., & Muglach, K. 2016, ApJ, 818, 203
Zhou, G., Wang, J., & Jin, C. 2013, SoPh, 283, 273